\documentclass{article}
\usepackage{ijcai23}

\usepackage{times}
\usepackage{soul}
\usepackage{url}
\usepackage[hidelinks]{hyperref}
\usepackage[utf8]{inputenc}
\usepackage[small]{caption}
\usepackage{graphicx}
\usepackage{amsmath}
\usepackage{amsthm}
\usepackage{booktabs}
\usepackage{algorithm}
\usepackage[switch]{lineno}

\usepackage{bm}
\usepackage[noend]{algpseudocode}
\usepackage{amsfonts}
\usepackage{xspace}
\usepackage{makecell}
\usepackage{multirow}
\usepackage{array}
\usepackage{color}
\usepackage{enumitem}

\newcommand{\ie}{\emph{i.e.,}\xspace}

\newcommand{\baby}{PFedRec\xspace}

\title{Dual Personalization on Federated Recommendation}

\author{
Chunxu Zhang$^{1,2}$
\and
Guodong Long$^3$\and
Tianyi Zhou$^4$\and
Peng Yan$^3$\and
Zijian Zhang$^{1,2}$\and
Chengqi Zhang$^3$\And
Bo Yang$^{1,2}$\thanks{Corresponding author.}
\affiliations
$^1$Key Laboratory of Symbolic Computation and Knowledge Engineering of Ministry of Education, China\\
$^2$College of Computer Science and Technology, Jilin University, China\\
$^3$Australian Artificial Intelligence Institute, FEIT, University of Technology Sydney\\
$^4$Computer Science and UMIACS, University of Maryland
\emails
\{cxzhang19, zhangzj2114\}@mails.jlu.edu.cn,
\{guodong.long, Chengqi.Zhang\}@uts.edu.au,
zhou@umiacs.umd.edu,
yanpeng9008@hotmail.com,
ybo@jlu.edu.cn
}

\begin{document}

\maketitle

\begin{abstract}
    Federated recommendation is a new Internet service architecture that aims to provide privacy-preserving recommendation services in federated settings. Existing solutions are used to combine distributed recommendation algorithms and privacy-preserving mechanisms. Thus it inherently takes the form of heavyweight models at the server and hinders the deployment of on-device intelligent models to end-users. This paper proposes a novel Personalized Federated Recommendation (PFedRec) framework to learn many user-specific lightweight models to be deployed on smart devices rather than a heavyweight model on a server. Moreover, we propose a new dual personalization mechanism to effectively learn fine-grained personalization on both users and items. The overall learning process is formulated into a unified federated optimization framework. Specifically, unlike previous methods that share exactly the same item embeddings across users in a federated system, dual personalization allows mild finetuning of item embeddings for each user to generate user-specific views for item representations which can be integrated into existing federated recommendation methods to gain improvements immediately. Experiments on multiple benchmark datasets have demonstrated the effectiveness of PFedRec and the dual personalization mechanism. Moreover, we provide visualizations and in-depth analysis of the personalization techniques in item embedding, which shed novel insights on the design of recommender systems in federated settings. The code is available\footnote{\href{https://github.com/Zhangcx19/IJCAI-23-PFedRec}{https://github.com/Zhangcx19/IJCAI-23-PFedRec}}.
\end{abstract}

\section{Introduction}\label{introduction}
Federated recommendation is a new service architecture for Internet applications, and it aims to provide personalized recommendation service while preserving user privacy in the federated settings. Existing federated recommendation systems~\cite{muhammad2020fedfast,yi2021efficient,perifanis2022federated,wu2022federated} are usually to be an adaptation of distributed recommendation algorithms by embodying the data locality in federated setting and adding privacy-preserving algorithms with guaranteed protection. However, these implementations of federated recommendations still inherit the traditional service architecture, which is to deploy large-scale models at servers. Thus it is impractical and inconsistent with the newly raised on-device service architecture, which is to deploy a lightweight model on the device to provide service independently without frequently communicating with the server. Given the challenge of implementing data locality on devices in federated settings, the personalization mechanism needs to be reconsidered to better capture fine-grained personalization for end-users.

Personalization is the core component of implementing federated recommendation systems. Inherited from conventional recommendation algorithms, existing federated recommendation frameworks are usually composed of three modules: user embedding to preserve the user's profile, item embedding to maintain proximity relationships among items, and the score function to predict the user's preference or rating for a given item. They usually preserve user-specific personalization in the user embedding module while sharing consensus on item embeddings and score functions. 

This paper proposes a new \textbf{dual personalization} mechanism designed to capture fine-grained two-fold personal preferences for users in the federated recommendation system. Inspired by human beings' decision logic, we believe all modules in the recommendation framework should be used to preserve part of personalization rather than use user embedding only. For example, the score function is to mimic the user's personal decision logic that is natural to be diverse across clients. Furthermore, given an item set, different people may have different views on measuring their proximity relationships. Therefore, personalized item embedding could be essential to capture people's personal preferences further. 

\begin{figure}[!t]
\centering
\includegraphics[width=0.5\textwidth,height=0.21\textwidth]{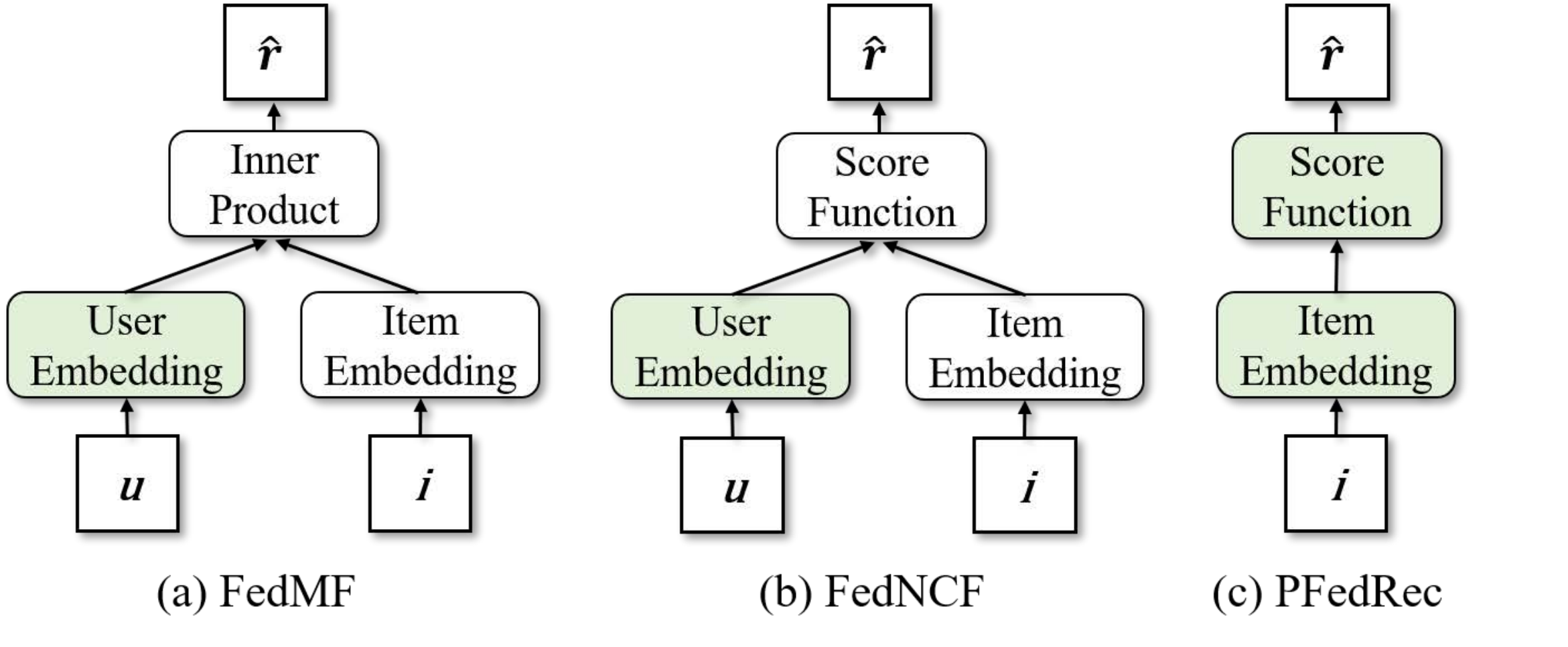}
\caption{Different frameworks for the personalized federated recommendation. The {\color{green} green block} represents a \textbf{personalized module}, which indicates the part of model is to preserve user preference. Our proposed model will preserve dual personalization on two modules.}
\label{fig:architecture}
\end{figure}

To implement the aforementioned ideas in federated settings, we propose a new federated recommendation framework to implement fine-grained personalization on multiple modules which are illustrated in Figure \ref{fig:architecture} (c). First, we use a personalized score function to capture user's preferences, and it could be implemented with a multi-layer neural networks. Second, we remove the user embedding from the federated recommendation framework because the current neural-based personalized score function has enough representation capability to preserve the information of user embeddings. Third, we implement light finetuning to learn personalized item embeddings in federated settings. This proposed decentralized intelligence architecture is a natural simulation of human beings' decision-making that each person has a relatively independent mind to make decisions. 

The learning procedure is also carefully tailored in a federated setting. A personalized score function will be learned using the user's own data on the device, and then it won't be sent to the server for global aggregation that usually generates a general view for all devices.
Moreover, the personalized item embedding will be implemented through light finetuning in a federated learning framework, thus it can leverage both the general view from server and the personalized view from user's own data. 

In summary, we propose a novel federated recommendation framework that integrates both the personalized score function and personalized item embedding via light finetuning from the shared item embedding. 
\textbf{Our key contributions} are summarized as follows. 
\begin{itemize}
    \item We propose a novel federated recommendation framework which is more naturally consistent with layer-wise neural architecture and can better fit federated learning. 
    \item We design a novel dual personalization mechanism to capture user preferences using a personalized score function and fine-grained personalization on item embeddings. It can be integrated with other baselines to improve their performances.
    \item We formulate the proposed federated recommendation learning problem into a unified federated optimization framework with a bi-level objective.
    \item Our method can significantly outperform existing federated recommendation baselines.  
\end{itemize}

\section{Related Work}\label{related work}
\subsection{Personalized Federated Learning} \label{sec:PFL}
\textbf{Federated Learning (FL)} is a new machine learning paradigm that a server orchestrates a large number of clients to train a model without accessing their data~\cite{kairouz2021advances,li2020federated,long2021federated,long2020federated,NEURIPS2022_7aa320d2,tan2022fedproto,chen2023prompt}.
The vanilla federated learning method, FedAvg~\cite{mcmahan2017communication}, is to learn a robust model at the server while embodying data locality for each device with non-IID data. \textbf{Personalized Federated Learning (PFL)} is to learn a personalized model for each device to tackle the non-IID challenge.  
Per-FedAvg~\cite{fallah2020personalized} exploits Model-Agnostic Meta-Learning (MAML)~\cite{finn2017model} to find a shared initial model for all clients and each client can learn a personalized model with its own data. Ditto~\cite{li2021ditto} proposes a bi-level optimization framework for PFL which introduces a regularization term by constraining the distance between the local and the global model. ~\cite{shamsian2021personalized} propose to replace the global model with a global hyper-network orchestrate clients' local training. FedRecon~\cite{singhal2021federated} is a meta-learning-based method that preserves a local model for each client and trains a global model collaboratively with FedAvg. There are also some attempts about graph-based~\cite{tan2022federated,ijcai2022p357} and cluster-based~\cite{ma2022convergence,long2023multi} methods. In this paper, we focus on developing personalization in the federated recommendation scenario where device data distributions are heavily non-IID, and it has not been well explored.

\subsection{Federated Recommendation Systems}
Federated recommendation has attracted much attention recently due to the rising privacy concern. Some recent works focus on only using the interaction matrix which is the most fundamental recommendation scenario. 
FCF~\cite{ammad2019federated} is the first FL-based collaborative filtering method, which employs the stochastic gradient approach to update the local model, and FedAvg is adopted to update the global model. Improving user privacy protection, FedMF~\cite{chai2020secure} adapts distributed matrix factorization to the FL setting and introduces the homomorphic encryption technique on gradients before uploading to the server. MetaMF~\cite{lin2020meta} is a distributed matrix factorization framework where a meta-network is adopted to generate the rating prediction model and private item embedding. ~\cite{wu2022federated} presents FedPerGNN where each user maintains a GNN model to incorporate high-order user-item information. However, the server in both MetaMF and FedPerGNN preserves all the recommendation model parameters which can be used to infer the user's interaction information, resulting in the risk of user privacy leakage. FedNCF~\cite{perifanis2022federated} adapts Neural Collaborative Filtering (NCF)~\cite{he2017neural} to the federated setting which introduces neural network to learn user-item interaction function to enhance model learning ability. 

Besides, there are federated recommendation methods using rich information that consider multiple data sources in modeling. FedFast~\cite{muhammad2020fedfast} extends FedAvg~\cite{mcmahan2017communication} with an active aggregation method to facilitate the convergence. Efficient-FedRec~\cite{yi2021efficient} decomposes the model into a news model on the server and a user model on the client, and reduces the computation and communication cost for users. Both works rely on more data sources, such as user features or news attributes rather than an interaction matrix.  ~\cite{lin2020fedrec,du2021federated,minto2021stronger,lin2021fr} are endeavors that focus on enhancing privacy of FedRec.
There are also attempts for other applications in FedRec, such as federated attack~\cite{10.1145/3534678.3539119,zhang2022pipattack}, social recommendation~\cite{liu2022federated}, Click-Through Rate (CTR) prediction~\cite{wu2022fedctr} and fair recommendation~\cite{liu2022fairness}. Existing federated recommendation methods usually combine the distributed recommendation algorithms and privacy-preserving techniques directly. They inherits the heavyweight models at the server and pay little attention on lightweight models deployed at client. In this paper, we present a novel federated recommendation framework aiming to learn many user-specific lightweight models rather than a heavyweight model on the server.

\section{Problem Formulation}
\textbf{Federated Learning} is to learn a global model parameterized by $\theta$ to serve all clients whose data are private. The optimal solution should minimize the accumulated loss of all clients,
\begin{equation} \label{eq:pf-FL}
    \min_\theta \sum_{i=1}^{N}\alpha_i L_i(\theta)
\end{equation}
where $L_i(\theta)$ is the supervised loss on the $i$-th client with dataset $D_i$, and all clients share the global parameter $\theta$. The $\alpha_i$ is a weight for the loss of the $i$-th client. For example, the conventional FL algorithm, FedAvg ~\cite{mcmahan2017communication}, defines $\alpha_i$ as the fraction of the size of the client's training data, \ie $\alpha_i := |D_i|/\sum_{j=1}^{N} |D_j|$. Once the global model is trained, it can be used for prediction tasks on all clients.

\textbf{Personalized Federated Learning} simultaneously leverages common knowledge among clients and learns a personalized model for each client, with learning objective as,
\begin{equation} \label{eq:pf-PFL}
  \begin{aligned} 
    \min_{\theta, \{\theta_i\}_{i=1}^{N}} \sum_{i=1}^{N}\alpha_iL_i(\theta,\theta_i)
  \end{aligned}
\end{equation}
where each client has a unique personalized parameter $\theta_i$, and $\theta$ is the global parameter as mentioned in Eq. (\ref{eq:pf-FL}). For example,~\cite{fallah2020personalized} leverage $\theta$ as initialization of $\theta_i$, \ie $\theta_i:=\theta-\nabla l_i(\theta)$, where $l_i(\theta)$ is the loss of a vanilla model on the $i$-th client. The $L_i(\theta, \theta_i)$ is then formulated as
\begin{equation} \label{eq:pf-meta}
    \begin{aligned}
    L_{i}(\theta,\theta_i):= l_i(\theta-\nabla l_i(\theta))
    \end{aligned}
\end{equation}

\textbf{Recommendation with Neural Networks} This work focuses on the fundamental scenario where recommendation only relies on the user-item interaction matrix without extra user/item attributes. The recommendation framework can be divided into three components: a user embedding module $\mathcal{E}$, an item embedding module $E$ and a score function $S$. We denote these modules' parameters with $\theta:=(\theta^u, \theta^m, \theta^s)$ and formulate the learning objective as,
\begin{equation} \label{eq:pf-Rec}
    \min_{\theta} L(\theta;r, \hat{r}) := \min_\theta L(\theta;r, S(\mathcal{E}(e^u), E(e^m)))
\end{equation}
where $e^u$ and $e^m$ are one-hot encodings representing users and items. $r$ is a user's rating to the given item and $\hat{r}$ is a prediction from the score function $S(\mathcal{E}(e^u), E(e^m))$. $L$ is the loss evaluation metric, which could be a \textbf{\textit{point-wise loss}} as used in~\cite{wang2016scalable,he2017neural}, or a \textbf{\textit{pair-wise loss}} as in~\cite{rendle2012bpr,wang2019neural}. It is worth noting that conventional Matrix Factorization (MF) methods could be viewed as a special case of the framework in Eq. (\ref{eq:pf-Rec}), \ie the conventional MF is a model where the score function $S$ is simplified as the inner product operator without learnable parameters, and the embedding of user/item is obtained by the decomposition of the user-item interaction matrix.

\section{Methodology}
In this section, we propose a novel \textbf{P}ersonalized \textbf{Fed}erated \textbf{Rec}ommendation (\baby) framework, which aims to simultaneously learn many user-specific recommendation models deployed on end devices.

\subsection{Objective Function} \label{sec:objective}
\paragraph{Federated Learning Objective} We regard each user as a client under FL settings. The on-device recommendation task is then depicted as a PFL problem. Particularly, the item embedding module $E_i$ is assigned to be a global component which learns common item information and the score function $S_i$ is maintained locally to learn personalized decision logic. To further capture the difference between users and achieve a preference-preserving item embedding, we devise a bi-level optimization objective,
\begin{equation}
\label{eq:pf-global-loss}
    \begin{aligned} 
    &\min_{\theta^m,\{\theta_i\}_{i=1}^{N}} &\sum_{i=1}^{N} \alpha_{i}L_i(\theta_i;r,\hat{r})\\
    &s.t. & \theta_i:=(\theta^m-\nabla_{\theta^m} L_i, \theta^s_i)
  \end{aligned}
\end{equation}
where $\theta_{i}:=(\theta^{m}_i,\theta^{s}_i)$ is the personalized parameter for $E_i$ and $S_i$, and $L_i$ will be evaluated on the $i$-th client local data $D_i$. Under this framework, \baby first tunes $E$ into a personalized item embedding module $E_i$, and then learns a lightweight local score function $S_i$ to make personalized predictions. Different from the conventional recommendation algorithms, the user embedding module $\mathcal{E}$ is depreciated since the personalization procedure on a client will automatically capture the client's preference. There is no use to learn extra embeddings to describe clients.

\paragraph{Loss for Recommendation}
Equipped with the item embedding module and score function, we formulate the prediction of $j$-th item by $i$-th user's recommendation model as,
\begin{equation}
    \hat{r}_{ij} = S_i(E_i(e^j))
\end{equation}
Particularly, we discuss the typical recommendation task with implicit feedback, that is, $r_{ij}=1$ if $i$-th user interacted with $j$-th item; otherwise $r_{ij}=0$. With the binary-value nature of implicit feedback, we define the loss function of $i$-th user as the \textit{binary cross-entropy loss},
\begin{equation} \label{eq:local-loss}
    L_i(\theta_i;r,\hat{r}) = - \sum_{(i,j) \in D_i} \log \hat r_{ij} - \sum_{(i,j') \in D_i^-} \log (1 - \hat r_{ij'})
\end{equation}
where $D_i^-$ is the negative instances set of user $i$. Notably, other loss functions can also be used, and here we choose the binary cross-entropy loss to simplify the description. Particularly, to construct $D_i^-$ efficiently, we first count all the uninteracted items set as,
\begin{equation}\label{eq:uninteracted-item}
    \mathcal{I}_i^- = \mathcal{I}\backslash
\mathcal{I}_i
\end{equation}
where $\mathcal{I}$ denotes the full item list and $\mathcal{I}_i$ is the interacted item set of $i$-th user. Then, we uniformly sample negative instances from $\mathcal{I}_i^-$ by setting the sampling ratio according to the number of observed interactions and obtain $D_i^-$.

\subsection{Dual Personalization}
We present a dual personalization mechanism to enable the proposed framework can preserve fine-grained personalization for both user and item. 

\textbf{Using partial-based federated model aggregation to learn personalized user score function on each device.}
Our proposed model is composed of a neural-based score function parameterized by $\theta^s$ and an item embedding module parameterized by $\theta^m$. The coordinator/server of federated system will iteratively aggregate model parameters or gradients collected from each participant/device. Due to the concern of personalization and privacy, we implement a partial model aggregation strategy by keeping the score function as a private module on devices while sharing the item embedding to the server. Therefore, the server only aggregates the gradients or parameters $\theta^m$ from the item embedding module. The user's personalized score function module $\theta^s$ won't be sent to the server and thus won't be aggregated. Generally, the simple and swift multi-layer neural network is capable of tackling most scenarios, which is convenient for client deployment.

\textbf{Finetuning the item embedding module to generate personalized representations for items on each device.} According to Eq. (\ref{eq:pf-global-loss}), the learning objective of $\theta^m$ could be viewed as searching for a ``good initialization" that could be fast adaptive to the learning task on different devices. It shares similar ideas with meta-learning-based methods~\cite{fallah2020personalized} which have a local loss in Eq. (\ref{eq:pf-meta}). However, our proposed method takes a different optimization strategy we call \textit{post-tuning}. Specifically, rather than directly tuning a global model on clients' local data, it first learns the local score function with the global item embedding, and then replaces the global item embedding with personalized item embedding obtained by finetuning $\theta^{m}$.

\subsection{Algorithm}
\paragraph{Optimization} To solve the optimization problem as described in Sec. \ref{sec:objective} - objective function, we conduct an alternative optimization algorithm to train the model. As illustrated in Algorithm \ref{algorithm}, when the client receives the item embedding from server, it first replaces its embedding with the global one, and then updates the score function while keeping the item embedding module fixed. Then the client updates the item embedding based on the updated personalized score function. Finally, the updated item embedding would be uploaded to the server for global aggregation.

\paragraph{Workflow} The overall algorithm workflow could be summarized into several steps as follows. The server is responsible for updating shared parameters and organizing all clients to complete collaborative training. At the beginning of federated optimization, the server initializes the model parameters, which would be used as initial parameters for all client models. In each round, the server selects a random set of clients and distributes the global item embedding $\theta^m$ to them. When local training is over, the server collects the updated item embedding from each client to perform global aggregation. We build on the simplified version of FedAvg, a direct average of locally uploaded item embeddings. The overall procedure is summarized in Algorithm \ref{algorithm}.

\begin{algorithm}[!t]
\caption{Dual Personalization for Federated Recommendation}
\hspace*{0.02in} {\bf ServerExecute:}
\begin{algorithmic}[1]
\State Initialize item embedding $\theta^m$ and score function $\theta^s$
\For{${t}=1, 2, ...$} \Comment{Global communication rounds}
\State $S_t \leftarrow$ (select a client set of size $n$ randomly from all $N$ clients)
\For{client $i \in S_t$ \textbf{in parallel}}
\State $\theta^m_i \leftarrow$ ClientUpdate($i, \theta^m$) \Comment{Distribute global item embedding to client for update}
\EndFor
\State $\theta^m \leftarrow \frac{1}{n} \sum_{i=1}^n \theta^m_i$ \Comment{Global aggregation over $n$ local updated item embeddings}
\EndFor
\end{algorithmic}
\hspace*{0.02in} {\bf ClientUpdate($\bm{i,\theta^m}$):}
\begin{algorithmic}[1]
\State Initialize $\theta^m_i$ with $\theta^m$
\State Initialize $\theta^s_i$ with the latest update
\State Count all uninteracted items set $\mathcal{I}_i^-$ with Eq. (\ref{eq:uninteracted-item})
\State Sample negative instances set $D_i^-$ from $\mathcal{I}_i^-$
\State $\mathcal{B} \leftarrow$ (split $D_i \cup D_i^-$ into batches of size $B$)
\For{$e$ from $1$ to $E$} \Comment{Local training epochs}
\For{batch $b \in \mathcal{B}$}
\State Compute $L_i(\theta_i;r,\hat{r})$ with Eq. (\ref{eq:local-loss}) \Comment{Model loss of batch data $b$}
\State $\theta^s_i \leftarrow \theta^s_i - \eta \nabla_{\theta^s_i} L_i$ \Comment{Score funtion update}
\State Compute $L_i(\theta_i;r,\hat{r})$ with Eq. (\ref{eq:local-loss}) \Comment{Model loss with the updated $\theta^s_i$}
\State $\theta^m_i \leftarrow \theta^m_i - \eta' \nabla_{\theta^m_i} L_i$ \Comment{post-tuning for personalized item embedding module}
\EndFor
\EndFor
\State {\bf Return} $\theta^m_i$ to server
\end{algorithmic}
\label{algorithm}
\end{algorithm}

\paragraph{Efficient on-device update} Focusing on the fundamental recommendation scenario, \ie with only user-item interaction matrix, the item embedding module $E_i$ dominates the parameter volume in the recommendation model due to large item set size, which brings challenges to end devices with limited computing resources. Generally, the items set that each user interacts with is much smaller than the complete item collection. Based on this observation, we propose that each device only needs to maintain the interacted positive items and sampled negative samples instead of the complete item embedding module, resulting in an efficient on-device update. For clarity, we continue to use $\theta^m$ in the Algorithm formulation. In practice, each device only needs to maintain a subset of the complete item embeddings.

\section{Discussions}
\subsection{Privacy on Federated Recommendation}
Privacy-preserving is an essential motivation to advance existing cloud-centric recommendation to client-centric recommendation service architecture. In general, the federated learning's decentralized framework can embody data locality and information minimization rules (GDPR) that could greatly mitigate the risk of privacy leakage~\cite{kairouz2019advances}. To provide service with privacy guarantee, the FL framework should be integrated with other privacy-preserving methods, such as Differential Privacy and secure communication. Our proposed framework derives the same decentralized framework from vanilla FL to preserve data locality. For example, to tackle the privacy leakage risk caused by sending item embedding to the server, we could simply apply differential privacy to inject noise into the embeddings so that the server cannot simply infer the updated items by watching the changes of embeddings. More analysis and experimental verification can be found in Sec. \ref{ldp}.

\subsection{A General Framework for Federated Recommendation}
The proposed framework in Figure \ref{fig:architecture} (c) could be a general form of federated recommendation because our framework could be easily transformed into an equivalent form of other frameworks. For example, if we assign the score function as a one-layer linear neural network, \baby is equal to FedMF in Figure \ref{fig:architecture} (a). Moreover, if we change the personalized score function from full personalization to partial layer personalization, our method could be equivalent to FedNCF in Figure \ref{fig:architecture} (b) which has a shared score function across clients. Furthermore, our proposed framework's architecture could be naturally aligned with the classic neural network architecture, thus it has a bigger potential to achieve better learning efficiency and is more flexible to extend.

\section{Experiments}
\subsection{Experimental Setup}
We evaluate the proposed \baby on four real-world datasets: MovieLens-100K, MovieLens-1M~\cite{harper2015movielens}, Lastfm-2K~\cite{Cantador:RecSys2011} and Amazon-Video~\cite{ni2019justifying}. They are all widely used datasets in assessing recommendation models. Specifically, two MovieLens datasets were collected through the MovieLens website, containing movie ratings and each user has at least 20 ratings. Lastfm-2K is a music recommendation dataset, and each user maintains a list of her favorite artists and corresponding tags. Amazon-Video was collected from the Amazon site, containing product reviews and metadata information. We excluded users with less than 5 interactions in Lastfm-2K and Amazon-Video. The characteristics of datasets are shown in Table~\ref{datasets}. For dataset split, We follow the prevalent leave-one-out evaluation~\cite{he2017neural}. We evaluate the model performance with Hit Ratio (HR) and Normalized Discounted Cumulative Gain (NDCG) metrics.
\begin{table}[!t]
\renewcommand\arraystretch{1.}
\centering
\small
\begin{tabular}{p{65pt}p{30pt}<{\centering}p{30pt}<{\centering}p{27pt}<{\centering}p{30pt}<{\centering}}
\hline
\pmb{Dataset} & \pmb{Interactions} & \pmb{Users} & \pmb{Items} & \pmb{Sparsity} \\
\hline
MovieLens-100K & 100,000 & 943 & 1,682 & 93.70\% \\
MovieLens-1M & 1,000,209 & 6,040 & 3,706 & 95.53\% \\
Lastfm-2K & 185,650 & 1,600 & 12,454 & 99.07\% \\
Amazon-Video & 63,836 & 8,072 & 11,830 & 99.93\% \\
\hline
\end{tabular}
\caption{Dataset statistics.}
\label{datasets}
\end{table}

\subsection{Baselines and Implementation Details}
\paragraph{Baselines}
Our method is compared with baselines in both centralized and federated settings. Focusing on the performance improvement of the infrastructure of recommendation models that all others derive from, we select the general and fundamental baselines that conduct recommendations only based on the interaction matrix.  
\begin{itemize}[leftmargin=*]
\item \textbf{Matrix Factorization (MF)}~\cite{koren2009matrix}: This method is a typical recommendation algorithm. Particularly, it decomposes the rating matrix into two embeddings located in the same latent space to characterize users and items, respectively.

\item \textbf{Neural Collaborative Filtering (NCF)}~\cite{he2017neural}: This method models user-item interaction function with an MLP, and is one of the most representative neural recommendation models. Specifically, we apply the interaction function with a three-layer MLP for comparison, which is adopted in the original paper.

\item \textbf{FedMF}~\cite{chai2020secure}: It is a federated version of MF which is a typical FedRec method. It updates user embedding locally and aggregates item gradients globally.

\item \textbf{FedNCF}~\cite{perifanis2022federated}: It is a federated version of NCF. Specifically, each user updates user embedding locally and uploads item embedding and score function to the server for global update.

\item \textbf{Federated Reconstruction (FedRecon)}~\cite{singhal2021federated}: It is a state-of-the-art PFL framework, and we test it under the matrix factorization scenario. Between every two rounds, this method does not inherit user embedding from the previous round but trains it from scratch.

\item \textbf{Meta Matrix Factorization (MetaMF)}~\cite{lin2020meta}: It is a distributed matrix factorization framework where a meta-network is adopted to generate the rating prediction module and private item embedding.

\item \textbf{Federated Graph Neural Network (FedPerGNN)~\cite{wu2022federated}}: It deploys a GNN in each client and the user can incorporate high-order user-item information by a graph expansion protocol.

\end{itemize}  

\paragraph{Implementation Details}
We sample 4 negative instances for each positive instance following ~\cite{he2017neural}. For all methods, we set the user (item) embedding size as 32 and the batch size is fixed as 256. For our method, we assign the score function with a one-layer MLP for simplification, which can be regarded as an enhanced FedMF with our dual personalization mechanism. We implement the methods based on the Pytorch framework\footnote{\href{https://github.com/Zhangcx19/IJCAI-23-PFedRec}{Code: https://github.com/Zhangcx19/IJCAI-23-PFedRec}} and run all the experiments for 5 repetitions and report the average results.
\begin{table*}[!t]
\centering
\scriptsize
\begin{tabular}{p{20pt}p{48pt}p{40pt}<{\centering}p{40pt}<{\centering}p{40pt}<{\centering}p{40pt}<{\centering}p{40pt}<{\centering}p{40pt}<{\centering}p{40pt}<{\centering}p{40pt}<{\centering}}
\hline 
& \multirow{2}{*}{\textbf{Method}} & \multicolumn{2}{c}{\textbf{MovieLens-100K}} &
\multicolumn{2}{c}{\textbf{MovieLens-1M}} &
\multicolumn{2}{c}{\textbf{Lastfm-2K}} &
\multicolumn{2}{c}{\textbf{Amazon-Video}} \\
& & HR@10 & NDCG@10 & HR@10 & NDCG@10 & HR@10 & NDCG@10 & HR@10 & NDCG@10 \\
\hline
\multirow{2}{*}{\textbf{CenRec}} &
\textbf{NCF} & 64.14 $\pm$ 0.98
 & 37.91 $\pm$ 0.37 & 64.17 $\pm$ 0.99 & 37.85 $\pm$ 0.68 & 82.44 $\pm$ 0.42 & 67.43 $\pm$ 0.89 & 60.16 $\pm$ 0.43 & 38.97 $\pm$ 0.14  \\
& \textbf{MF} & 64.43 $\pm$ 1.02 & 38.95 $\pm$ 0.56 & 68.45 $\pm$ 0.34 & 41.37 $\pm$ 0.18 & 82.71 $\pm$ 0.54 & 71.04 $\pm$ 0.62 & 46.69 $\pm$ 0.65 & 29.83 $\pm$ 0.45\\
\cline{2-10}
\multirow{6}{*}{\textbf{FedRec}} &
\textbf{FedMF} & 65.15 $\pm$ 1.16 & 39.38 $\pm$ 1.08 & 67.72 $\pm$ 0.14 & 40.90 $\pm$ 0.14 & 81.64 $\pm$ 0.48 & 69.36 $\pm$ 0.42 & 59.67 $\pm$ 0.19 & 38.55 $\pm$ 0.21 \\
& \textbf{FedNCF} & 60.62 $\pm$ 0.59 & 33.25 $\pm$ 1.35 & 60.54 $\pm$ 0.46 & 34.17 $\pm$ 0.40 & 81.55 $\pm$ 0.38 & 61.03 $\pm$ 0.63 & 57.77 $\pm$ 0.07 & 36.86 $\pm$ 0.06 \\
& \textbf{FedRecon} & 64.45 $\pm$ 0.81 & 37.78 $\pm$ 0.38 & 63.28 $\pm$ 0.15 & 36.59 $\pm$ 0.33 & 82.06 $\pm$ 0.38 & 67.58 $\pm$ 0.35 & 59.80 $\pm$ 0.14 & 38.87 $\pm$ 0.13 \\
& \textbf{MetaMF} & 66.38 $\pm$ 0.24 & 40.59 $\pm$ 0.31 & 45.61 $\pm$ 0.18 & 25.24 $\pm$ 0.35 & 80.88 $\pm$ 0.45 & 64.24 $\pm$ 0.45 & 57.51 $\pm$ 0.53 & 37.25 $\pm$ 0.28 \\
& \textbf{FedPerGNN} & 10.50 $\pm$ 0.12 & 4.92 $\pm$ 0.21 & 9.69 $\pm$ 0.23 & 4.37 $\pm$ 0.31 & 10.19 $\pm$ 0.41 & 4.83 $\pm$ 0.25 & 10.72 $\pm$ 0.33 & 4.90 $\pm$ 0.32 \\
& \textbf{\baby (Ours)} & \textbf{71.62} $\pm$ \textbf{0.83} & \textbf{43.44} $\pm$ \textbf{0.89} & \textbf{73.26} $\pm$ \textbf{0.20} & \textbf{44.36} $\pm$ \textbf{0.16} & \textbf{82.38} $\pm$ \textbf{0.92} & \textbf{73.19} $\pm$ \textbf{0.38} & \textbf{60.08} $\pm$ \textbf{0.08} & \textbf{39.12} $\pm$ \textbf{0.09} \\
\hline
\end{tabular}
\caption{Performance of HR@10 and NDCG@10 on four datasets. \textbf{CenRec} and \textbf{FedRec} represent centralized and federated methods, respectively. The results are the mean and standard deviation of five repeated trials.}
\label{performance comparison}
\end{table*}

\begin{table*}[!ht]
\centering
\scriptsize
\begin{tabular}{p{40pt}p{40pt}<{\centering}p{40pt}<{\centering}p{40pt}<{\centering}p{40pt}<{\centering}p{40pt}<{\centering}p{40pt}<{\centering}p{40pt}<{\centering}p{40pt}<{\centering}}
\hline 
\multirow{2}{*}{\textbf{Method}} & \multicolumn{2}{c}{\textbf{MovieLens-100K}} &
\multicolumn{2}{c}{\textbf{MovieLens-1M}} &
\multicolumn{2}{c}{\textbf{Lastfm-2K}} &
\multicolumn{2}{c}{\textbf{Amazon-Video}} \\
& HR@10 & NDCG@10 & HR@10 & NDCG@10 & HR@10 & NDCG@10 & HR@10 & NDCG@10 \\
\hline
\textbf{FedMF} & 65.15 $\pm$ 1.16 & 39.38 $\pm$ 1.08 & 67.72 $\pm$ 0.14 & 40.90 $\pm$ 0.14 & 81.64 $\pm$ 0.48 & 69.36 $\pm$ 0.42 & 59.67 $\pm$ 0.19 & 38.55 $\pm$ 0.21 \\
\textbf{w/ DualPer} & 71.62 $\pm$ 0.83 & 43.44 $\pm$ 0.89 & 73.26 $\pm$ 0.20 & 44.36 $\pm$ 0.16 & 82.38 $\pm$ 0.92 & 73.19 $\pm$ 0.38 & 60.08 $\pm$ 0.08 & 39.12 $\pm$ 0.09 \\
\textbf{Improvement} & $\uparrow \textbf{9.93\%}$ & $\uparrow \textbf{10.31\%}$ & $\uparrow \textbf{8.18\%}$ & $\uparrow \textbf{8.46\%}$ & $\uparrow 0.91\%$ & $\uparrow \textbf{5.52\%}$ & $\uparrow 0.69\%$ & $\uparrow 1.48\%$ \\
\hline
\textbf{FedNCF} & 60.62 $\pm$ 0.59 & 33.25 $\pm$ 1.35 & 60.54 $\pm$ 0.46 & 34.17 $\pm$ 0.40 & 81.55 $\pm$ 0.38 & 61.03 $\pm$ 0.63 & 57.77 $\pm$ 0.07 & 36.86 $\pm$ 0.06 \\
\textbf{w/ DualPer} & 68.82 $\pm$ 1.35 & 39.33 $\pm$ 0.85 & 68.17 $\pm$ 0.55 & 39.56 $\pm$ 0.29 & 82.31 $\pm$ 0.56 & 71.64 $\pm$ 0.43 & 59.57 $\pm$ 0.57 & 38.73 $\pm$ 0.62 \\
\textbf{Improvement} & $\uparrow \textbf{13.53\%}$ & $\uparrow \textbf{18.29\%}$ & $\uparrow \textbf{12.60\%}$ & $\uparrow \textbf{15.77\%}$ & $\uparrow 0.93\%$ & $\uparrow \textbf{17.38\%}$ & $\uparrow 3.12\%$ & $\uparrow \textbf{5.07\%}$ \\
\hline
\textbf{FedRecon} & 64.45 $\pm$ 0.81 & 37.78 $\pm$ 0.38 & 63.28 $\pm$ 0.15 & 36.59 $\pm$ 0.33 & 82.06 $\pm$ 0.38 & 67.58 $\pm$ 0.35 & 59.80 $\pm$ 0.14 & 38.87 $\pm$ 0.13 \\
\textbf{w/ DualPer} & 70.20 $\pm$ 0.90 & 41.83 $\pm$ 0.71 & 68.89 $\pm$ 0.26 & 40.04 $\pm$ 0.16 & 83.51 $\pm$ 0.23 & 74.83 $\pm$ 0.44 & 60.23 $\pm$ 0.16 & 39.20 $\pm$ 0.12 \\
\textbf{Improvement} & $\uparrow \textbf{8.92\%}$ & $\uparrow \textbf{10.72\%}$ & $\uparrow \textbf{8.87\%}$ & $\uparrow \textbf{9.43\%}$ & $\uparrow 1.77\%$ & $\uparrow \textbf{10.73\%}$ & $\uparrow 0.72\%$ & $\uparrow 0.85\%$ \\
\hline
\end{tabular}
\caption{Performance improvement for integrating our dual personalization mechanism (\textbf{DualPer}) to three federated baseline algorithms. The results are the mean and standard deviation of five repeated trials, and the significant improvements (over 5\%) are highlighted.}
\label{integration}
\end{table*}

\subsection{Comparison Analysis}
We conduct experiments on four datasets for performance comparison and the resuls are shown in Table \ref{performance comparison}.

\paragraph{Results \& discussion}
From the results, we have several observations: (1) \textbf{\baby obtains better performance than centralized methods in some cases.} In the centralized scenario, only user embedding is regarded as the personalized component to learn user characteristics, and other components are totally shared among users. In comparison, our dual personalization mechanism considers two forms of personalization, which can further exploit user preferences. (2) \textbf{\baby realizes outstanding advances on the two MovieLens datasets.} In these two datasets, each user has more interaction samples which can be used to train device recommendation models, hence promoting user personalization learning and fitting our method better. (3) \textbf{\baby consistently achieves the best performance against all federated methods.} In FedRec, the common item embeddings help transfer the shared information among users, which facilitates collaborative training of individual user models.
However, different users present rather distinct preferences for items and existing federated methods deploy the global item embeddings indiscriminately for all clients ignoring user-specific preferences. In comparison, our dual personalization mechanism learns fine-grained personalization which fits user preferences.

\subsection{Enhance Federated Recommendation Methods with Our Dual Personalization Mechanism} 
This paper proposes a lightweight dual personalization mechanism to enhance personalization handling, which can be easily integrated into federated learning methods. Particularly, we take FedMF, FedNCF and FedRecon as examples to verify the efficacy of the dual personalization mechanism.

\paragraph{Results \& discussion}
According to Table \ref{integration}, all three federated recommendation methods are significantly improved by integrating our dual personalization mechanism. Among them, FedNCF attains the most remarkable boost. The highest HR@10 and NDCG@10 increase exist on MovieLens-100K, \ie 13.53\% and 18.29\%. Compared with Lastfm-2K and Amazon-Video, the improvement of the dual personalization mechanism is more evident on the two MovieLens datasets, almost around 10\%, where each user has more samples locally and facilitates user preference capture. In summary, our proposed dual personalization mechanism can help the local model to learn user-specific item embedding, which benefits the recommendation system prominently.

\subsection{A Close Look of Personalization in \baby}
To further verify and analyze the role of personalized item embedding in our method, we conduct empirical experiments to answer the following questions:
\begin{itemize}[leftmargin=*]
\item \textit{\textbf{Q1}: Why personalized item embeddings benefit recommendation more than the global one?}

\item \textit{\textbf{Q2}: How specific are the personalized item embeddings among users?}
\end{itemize}

\begin{figure*}[!t]
\centering
\includegraphics[width=1.\textwidth,height=0.24\textwidth]{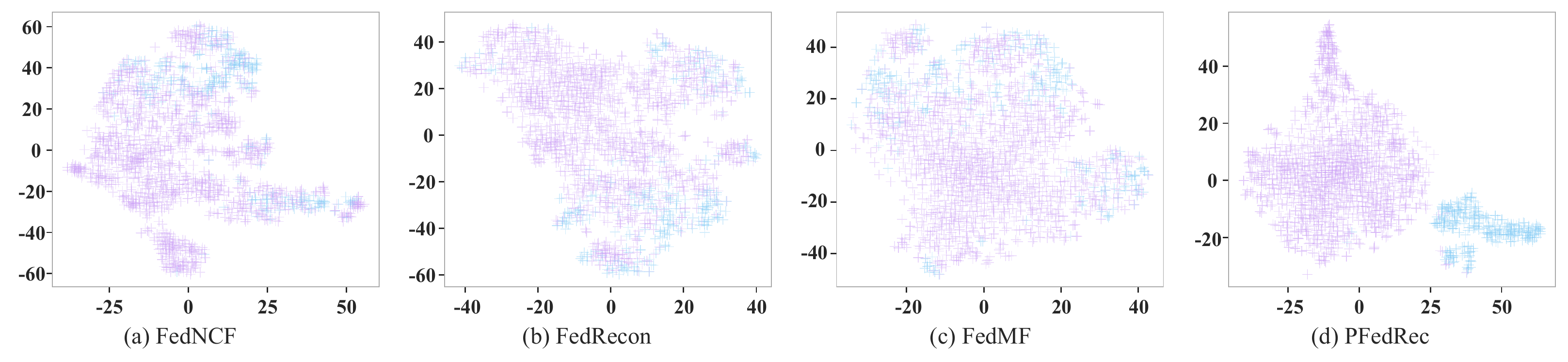}
\caption{TSNE visualization of item embeddings learned by baselines and our method.}
\label{Q1}
\end{figure*}
\textbf{To answer \textit{Q1}}, 
We first discuss its straightforward insight, then we present visualization to demonstrate our claim. The recommendation system aims to provide user-specific recommendations by exploiting historical interactions. In the FedRec setting, item embedding is consistently considered to maintain the common characteristics among users, and its role in depicting user-specific preferences has been neglected. On the other hand, describing users with common item embedding introduces noisy information, which may incur unsuitable recommendations. Through personalizing item embedding, we enhance personalization modeling in federated learning methods, which depicts the user-specific preference.

We compare the item embedding learned by baselines (top-3 FedRec methods for performance due to limited space) with our method. Particularly, we select a user randomly from the MovieLens-100K dataset and visualize the embeddings by mapping them into a 2-D space through t-SNE~\cite{maaten2008visualizing}. In this paper, we mainly focus on the implicit feedback recommendation, so each item is either a positive or negative sample of the user. 
As shown in Figure \ref{Q1}, the item embeddings of positive (blue) and negative (purple) samples are mixed in baselines, where all users share the global item embeddings. However, they can be divided into two clusters by \baby. We can easily conclude that our model learns which items the user prefers.

\begin{figure}[!t]
\centering
\includegraphics[width=0.48\textwidth,height=0.18\textwidth]{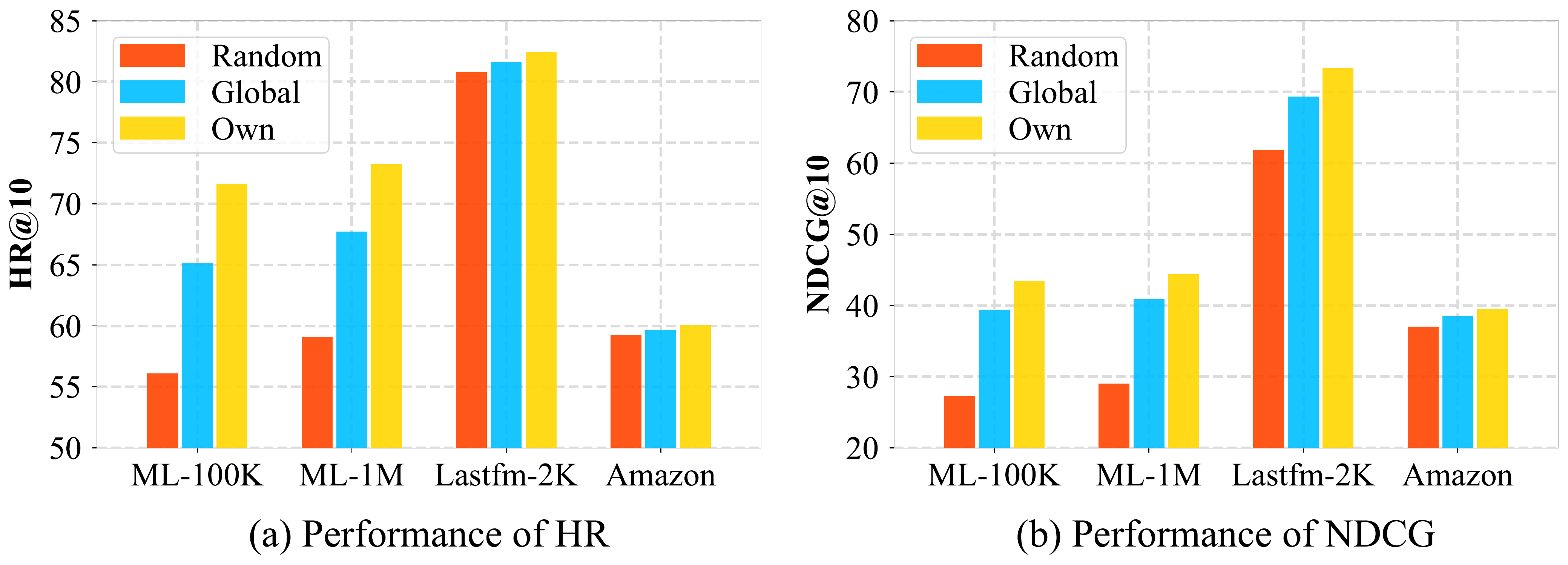}
\caption{Client inference using different item embeddings.}
\label{Q2}
\end{figure}
\textbf{To answer \textit{\textbf{Q2}}},
we compare three usages of item embedding during inference as follows: 

-- \textbf{\textit{Random}}: Each client is assigned with item embedding from a random user, \ie every client runs with its score function and item embedding from a random user.

-- \textbf{\textit{Global}}: We assign each client with globally shared item embedding, \ie every client runs with its score function and {global} item embedding.

-- \textbf{\textit{Own}}: It follows our setting that every client runs with its score function and personalized item embedding.

Specifically, we first train \baby, then assign the learned item embeddings as the above three ways for inference. As shown in Figure \ref{Q2}, clients with their item embedding achieve the best performance, and clients with item embedding from others degrade significantly. Item embedding from a random user contains little helpful information for inference, even less than the common characteristics in global item embedding. The personalized item embedding learned by \baby has been adapted to client preference, and different clients achieve rather distinct item embeddings, depicting the user-specific preference.

\subsection{Protection with Local Differential Privacy}\label{ldp}
To enhance the preservation of user privacy, we integrate the Local Differential Privacy (LDP) technique \cite{choi2018guaranteeing} into our framework. Particularly, we add the zero-mean Laplacian noise to the client’s item embedding before uploading to the server, \ie $\theta^m=\theta^m+Laplace(0,\lambda)$, and $\lambda$ is the noise strength. We set $\lambda=[0,0.1,0.2,0.3,0.4,0.5]$ to test our method's performance.
\begin{table}[!t]
\centering
\scriptsize
\begin{tabular}{p{25pt}p{43pt}p{13pt}<{\centering}p{13pt}<{\centering}p{13pt}<{\centering}p{13pt}<{\centering}p{13pt}<{\centering}p{13pt}<{\centering}}
\hline
\textbf{Dataset} & \textbf{Noise strength} & \textbf{$\bm \lambda$=0} & \textbf{$\bm \lambda$=0.1} & \textbf{$\bm \lambda$=0.2} & \textbf{$\bm \lambda$=0.3} & \textbf{$\bm \lambda$=0.4} & \textbf{$\bm \lambda$=0.5} \\
\hline
\multirow{2}*{\textbf{ML-100K}} & HR@10 & \bm{$71.62$} & $71.45$ & $71.26$ & $71.13$ & $70.84$ & $70.88$ \\
 & NDCG@10 & \bm{$43.44$} & $43.36$ & $43.30$ & $43.22$ & $43.14$ & $43.21$ \\
\hline
\multirow{2}*{\textbf{ML-1M}} & HR@10 & \bm{$73.26$} & $73.13$ & $73.19$ & $73.05$ & $73.18$ & $73.08$ \\
 & NDCG@10 & \bm{$44.36$} & $44.16$ & $44.25$ & $44.26$ & $44.23$ & $44.18$ \\
\hline
\multirow{2}*{\textbf{Lastfm-2K}} & HR@10 & \bm{$82.38$} & $82.04$ & $81.91$ & $81.85$ & $81.98$ & $81.88$ \\
 & NDCG@10 & \bm{$73.19$} & $72.41$ & $ 72.23$ & $72.43$ & $72.39$ & $72.36$ \\
\hline
\multirow{2}*{\textbf{Amazon}} & HR@10 & \bm{$60.08$} & $59.31$ & $59.29$ & $59.21$ & $59.15$ & $59.06$ \\
 & NDCG@10 & \bm{$39.12$} & $37.97$ & $37.92$ & $ 37.83$ & $37.81$ & $37.34$\\
\hline
\end{tabular}
\caption{Performance of integrating LDP into our method with various Laplacian noise strength $\lambda$.}
\label{dp}
\end{table}

As shown in Table \ref{dp}, performance declines slightly as the noise strength $\lambda$ grows, while the performance drop is still acceptable. For example, when we set $\lambda=0.4$, the performance is also better than baselines in most cases. Hence, a moderate noise strength is desirable to achieve a good balance between recommendation accuracy and privacy protection.

\section{Conclusion}
This paper proposes a novel personalized federated recommendation framework to learn many on-device models simultaneously. We are the first to design the dual personalization mechanism that can learn fine-grained personalization on both users and items. This work could be fundamental work to pave the way for implementing a new service architecture with better privacy preservation, fine-grained personalization, and on-device intelligence. Given the complex nature of modern recommendation applications, such as cold-start problems, dynamics, using auxiliary information, and processing multi-modality contents, our proposed framework is simple and flexible enough to be extended to handle many new challenges. Moreover, the proposed dual personalization is a simple-yet-effective mechanism to be easily integrated with existing federated recommendation systems.

\section*{Acknowledgements}

Chunxu Zhang and Bo Yang are supported by the National Key R\&D Program of China under Grant Nos. 2021ZD0112501 and 2021ZD0112502; the National Natural Science Foundation of China under Grant Nos. U22A2098, 62172185, 62206105 and 62202200; Jilin Province Capital Construction Fund Industry Technology Research and Development Project No. 2022C047-1; Changchun Key Scientific and Technological Research and Development Project under Grant No. 21ZGN30.

\bibliography{ijcai23}
\bibliographystyle{ijcai23}

\end{document}